\documentclass[a4paper,fleqn]{cas-dc}

\usepackage[numbers,sort&compress]{natbib}
\usepackage{subfigure}
\usepackage{bm}
\usepackage{ulem}
\usepackage{multirow}
\usepackage{array}

\def\tsc#1{\csdef{#1}{\textsc{\lowercase{#1}}\xspace}}
\tsc{WGM}
\tsc{QE}
\tsc{EP}
\tsc{PMS}
\tsc{BEC}
\tsc{DE}

\begin{document}
\let\WriteBookmarks\relax
\def\floatpagepagefraction{1}
\def\textpagefraction{.001}
\shortauthors{Qi Meng et~al.}

\title [mode = title]{Doubly heavy tetraquarks including one-pion exchange potential}

\author[1]{Qi Meng}
\ead{qimeng@nju.edu.cn} 

\author[2,3]{Emiko Hiyama}
\ead{hiyama@riken.jp}

\author[4,5]{Makoto Oka}
\ead{makoto.oka@riken.jp}

\author[6,5]{Atsushi Hosaka}
\ead{hosaka@rcnp.osaka-u.ac.jp}

\author[1]{Chang Xu}
\ead{cxu@nju.edu.cn}

\address[1]{Department of Physics, Nanjing University, Nanjing, 210093, China}
\address[2]{Department of Physics, Tohoku University, Sendai, 980-8578, Japan}
\address[3]{RIKEN, Nishina Center, Wako, Saitama, 351-0198, Japan}
\address[4]{Few-body Systems in Physics Laboratory, RIKEN Nishina Center, Wako 351-0198, Japan}
\address[5]{Advanced Science Research Center, Japan Atomic Energy Agency (JAEA), Tokai 319-1195, Japan}
\address[6]{Research Center for Nuclear Physics (RCNP), Ibaraki, Osaka 567-0047, Japan}


\begin{abstract}
Spectrum of the doubly heavy tetraquarks is studied in a constituent quark model including one-pion exchange (OPE) potential. 
Central and tensor forces induced by OPE between two light quarks are considered.
Our results show that $I(J^P)=0(1^+)$ compact bound states are shifted up because of the repulsive central force between $\bar{q}\bar{q}$. This effect possibly leads to the small binding energy in $T_{cc}$.
In addition, a $I(J^P)=1(1^+)$ resonant state is reported with $ E=10641 \ \rm{MeV},\Gamma=15 \ \rm{MeV}$ and $ E=10640 \ \rm{MeV},\Gamma=15 \ \rm{MeV}$, without and with including OPE potential, respectively.
The repulsive central force and attractive tensor force almost cancel with each other and leave a small energy difference when OPE potential is included.
\end{abstract}


\maketitle

\section{Introduction}
Observation of a double-heavy tetraquark state, $T_{cc}$ \cite{LHCb:2021vvq}, by the LHCB collaboration is interesting in many senses. 
It showed a sharp resonance peak just below the threshold of $D^{+*}D^0$ threshold 
in the invariant mass spectrum of $D^0D^0\pi^+$ channel at only 5.7 MeV above the threshold. 
The resonance contains two charm quarks and the minimal composites are four quarks, $cc\bar{u}\bar{d}$.
The similarity to the $X(3872)$ \cite{Belle:2003nnu} resonance is remarkable. 
The $X(3872)$ is the first candidate of tetraquark with hidden $c\bar{c}$ and 
appears just at the threshold of $D^0\bar{D}^{0*}$ as a sharp peak.
Its spin and parity quantum numbers are confirmed as $J^P = 1^+$, while isospin is almost maximally broken.
Such appearances of sharp peaks at the vicinity of a threshold of two hadrons are suspected not being accidental,
as there are some other cases, such as $P_c (=uudc\bar{c})$ \cite{LHCb:2015yax} resonances around at the $D^*\Sigma_c$ threshold.
The common feature among them is that the resonances are most likely exotic, containing more than three quarks,
and very narrow. They often appear for heavy quarks, $c$ or $b$.
Then a natural question is what, if any, is the mechanism to make a narrow resonance at a threshold of two heavy hadrons.

There have been various approaches to understand the origin and structure of $T_{cc}$ (and the other exotic resonances) \cite{Ballot:1983iv, Zouzou:1986qh, Carlson:1987hh, Silvestre-Brac:1993zem, Pepin:1996id, Gelman:2002wf, Vijande:2003ki, Ebert:2007rn, Yang:2009zzp, Ikeda:2013vwa, Luo:2017eub, Karliner:2017qjm, Eichten:2017ffp, Liu:2019stu, Junnarkar:2018twb, Francis:2018jyb, Lu:2020rog, Deng:2018kly, Cheng:2020wxa, Braaten:2020nwp, Dong:2021bvy, Chen:2021vhg, Weng:2021hje, Chen:2021cfl, Deng:2021gnb, Padmanath:2022cvl, Kim:2022mpa, Cheng:2022qcm}.
Two distinct pictures are available, a compact tetraquark or a loosely-bound hadronic molecule.
At the threshold the latter component is supposed to be dominant, while the compact state may become dominant for a deeper
bound or resonant state. In reality, they may coexist as a linear combination.
The probability of the molecular state is called compositeness, which 
is known to be related to the scattering length of the threshold hadrons for the resonances around the threshold
(weak binding regime).

Prediction of a compact tetraquark state for $T_{cc}$ was made long time ago in the quark model.
The color Coulomb interaction between $cc$ makes a deep S-wave bound state with spin 1. It  is combined with
the scalar $S=0$ and isoscalar $I=0$ $ud$ diquark, which is the most attractive diquark due to the color-magnetic 
interaction between quarks.
Then the ground state with orbital $S$ state will have $I=0$ and $J^{\pi}=1^+$.
On the other hand, the fall-apart decay threshold with $J^{\pi}=1^+$ will be $D+D^*$,%
\footnote{Besides the fall-apart decay, $T_{cc}(cc\bar{u}\bar{d}) \to D(c\bar{u})+D^{*}(c\bar{d})$, 
it has strong $DD\pi$ and electromagnetic $DD\gamma$ decay channels, 
if the mass is above the corresponding threshold. 
If the mass of $T_{cc}$ is below $DD$ threshold, it decays only by weak interaction.}
where no strong color Coulomb interaction between $cc$ and the color magnetic interaction for $D^*$ is repulsive.
Therefore even with a cost of kinetic energy, a compact tetraquark state may appear below or around the $DD^*$
threshold.
It is interesting to note that the predicted spin-parity quantum numbers coincide with those of $X(3872)$ resonance.

Such compact tetraquark states for heavier quarks, $T_{bb} (bb\bar{u}\bar{d})$
or $T_{cb}(cb\bar{u}\bar{d})$, are expected to be bound deeply in the quark model. 
Thus it is very important to determine the structure and the component of the $T_{cc}$ wave function to
predict the spectrum of $T_{bb}$ and $T_{bc}$ states.%
\footnote{The tetraquark component of $X(3872) \sim c\bar{c}d\bar{d}$ has a similar advantage to the $D\bar{D}^*$ threshold, 
but it can decay by fall-apart into $J/\psi(c\bar{c})+\rho/\omega(d\bar{d}) \to J/\psi+2\pi$ and its $Q$-value is
large $\sim 500$ MeV.
The corresponding heavier tetraquarks, $X_b (b\bar{b}q\bar{q}$) or $X_{bc} (b\bar{c}q\bar{q}$), will be lower lying.}

In view of the above situation, it is important to apply the current best quark model Hamiltonian to the general $QQ \bar{q}\bar{q}$ states and carry out the calculation treating both the compact and molecular states in the same footing.
Indeed, we need a calculation of tetraquark states with the full coupling to the fall-apart scattering states.
Such calculation is possible by the use of variational method for the general four-quark system.
We here report our calculation using the Gaussian Expansion  Method with the real scaling method to explore both the bound/resonant states and the meson-meson scattering states.
The quark model Hamiltonian is fine-tuned to the quarkonia and heavy meson spectrum. 
We further consider the pion exchange between the light quarks, which induces the pion exchange interaction between $D^{(*)}/B^{(*)}$ mesons.

In our previous paper, we carried out the four-body quark model calculation for $T_{bb}$ and $T_{cc}$
and revealed several bound and resonant states without considering the pion exchange force between
light quarks \cite{Meng:2020knc, Meng:2021yjr}.
Our goal in this paper is to determine the contributions of the pion exchange interaction to the compact tetraquarks
and the resonant states. We also study the contributions of higher angular momenta ($L>0$) and
the role of the tensor force induced by the pion exchange.

The contents of the paper is as follows. 
After the introduction, the Hamiltonian and the employed computational method are discussed in Sec.~\ref{Hami}. We discuss our results in Sec.~\ref{results} and give a summary in Sec.~\ref{summary}.

\

\section{Hamiltonian and method}\label{Hami}

The Hamiltonian for the four-quark system is given by
\begin{eqnarray}
\label{Hamiltonian}
\begin{aligned}
	H=&\sum_{i}^{4}\Big(m_i+\frac{{\boldsymbol{p}_i}^{2}}{2m_i}\Big)-T_G \\
	&-\frac{3}{16}\sum_{i<j=1}^{4}\sum_{a=1}^{8}(\lambda_i^a \cdot \lambda_j^a) V_{ij}(\boldsymbol{r}_{ij})+ V_{\bar{q}\bar{q}}^{C} + V_{\bar{q}\bar{q}}^{T},
\end{aligned}
\end{eqnarray}
where $m_i$ and $\boldsymbol{p}_i$ are the mass and momentum of the $i^{th}$ quark, respectively, 
and $T_G$ is the kinetic energy of the center-of-mass motion that must be subtracted for the calculation of tetraquark masses. 
$\lambda_i^a$ are the SU(3) Gell-Mann matrices in the color space with color index $a \ (= 1, \dots 8)$ acting on the $i^{th}$ quark.
The quark-quark potential $V_{ij}(\boldsymbol{r})$ is taken from the AP1 of Ref.\cite{SilvestreBrac:1996bg},
\begin{eqnarray}
\label{potential}
\begin{aligned}
	V_{ij}(\boldsymbol{r})=&-\frac{\kappa}{r}+\lambda r^p-C \\
	&+\frac{2\pi\kappa'}{3m_i m_j}\frac{\mathrm{exp}(-r^2/r_0^2)}{\pi^{3/2}r_0^3}({\bm\sigma}_i\cdot{\bm\sigma}_j),
\end{aligned}
\end{eqnarray}
with
\begin{eqnarray}
\label{r0}
\begin{aligned}
r_0(m_i,m_j)=A \Big( \frac{{2m_i m_j}}{{m_i+m_j}} \Big)^{-B}.
\end{aligned}
\end{eqnarray}
$V_{ij}(\boldsymbol{r})$ consists of the color Coulomb potential, the power-law confining part, a (color-electric) constant term and the color-magnetic spin-spin interaction term. 

In order to reproduce the masses of relevant threshold mesons accurately, especially $D^0 D^{*+} (3875.1 \ \rm MeV)$, the threshold closest to $T_{cc}$, 
we have tuned the parameters in the Hamiltonian. The parameters and the corresponding masses of the heavy mesons are listed in Tables \ref{parameters} and \ref{mesons}, respectively.
The masses of the $b\bar{b}$ and $b\bar{q}$ mesons are reproduced within errors less than 15 MeV. 
The errors for the masses of $c\bar{c}$ and $c\bar{q}$ mesons are smaller as less than 1 MeV and the calculated mass of $D^0 D^{*+}$ threshold is 3875.0 MeV.
\begin{center}
\linespread{1.3}
\begin{table}
\caption{The tuned parameters of the employed Hamiltonian.}\label{parameters}
\begin{tabular}{p{1.5cm}<{\centering} p{1.5cm}<{\centering} | p{1.5cm}<{\centering} p{1.5cm}<{\centering}}
\toprule
\multicolumn{4}{c}{parameters}\\
\midrule
$p$			   	&2/3			&$m_{u,d}$(GeV) 	& 0.298\\
$\kappa$ 		& 0.4222 	&$m_{u,d}$(GeV) 	& 1.829\\
$\kappa'$ 		& 1.7525 	&$m_{u,d}$(GeV) 	& 5.206\\
$\lambda$(GeV$^{p+1}$) & 0.3798  &$m_{\pi}$(GeV)& 0.138  \\
$C$(GeV) 	& 1.1404  		&$f_{\pi}$(GeV)		& 0.093\\
$B$ 			& 0.3263 	& $g_A$ & 1 \\
$A$(GeV$^{B-1}$)& 1.5296 	& $\Lambda$(GeV) & 1\\
\bottomrule
\end{tabular}
\end{table}
\end{center} 

\begin{center}
\linespread{1.3}
\begin{table}
\caption{Calculated masses (Cal.) of heavy mesons compared with their experimental values (Exp.).}\label{mesons}
\begin{tabular}{p{1.0cm}<{\centering} p{1.0cm}<{\centering}  p{1.8cm}<{\centering} p{1.8 cm}<{\centering}}
\toprule
       & $J^P$	& Cal.(MeV)  &  Exp.(MeV) \\
\midrule
	$\eta_c$   	&	$0^-$	&  2983.1  &  2983.9  \\
	$J/\psi$   	&	$1^-$	&  3097.8  &  3096.9  \\
	$\eta_b$   	&	$0^-$	&  9387.9  &  9398.7  \\
	$\Upsilon$  &	$1^-$	&  9445.0  &  9460.3  \\
	$D^0$   		&	$0^-$	&  1865.1  &  1864.8  \\
	$D^{*+}$   	&	$1^-$	&  2009.9  &  2010.2  \\
	$B^0$   		&	$0^-$	&  5279.8  &  5279.7  \\
	$B^{*}$  	&	$1^-$	&  5334.0  &  5324.7  \\
 	
\bottomrule
\end{tabular}
\end{table}
\end{center} 

In the molecular-type states, the pion is considered to play an important role in forming the bound/resonant states by mediating a long-range interaction between the component hadrons.
On the other hand, in the microscopic picture of tetra-quark system, we consider that the pion couples directly to the light quarks and induces the effective long-range potential between the heavy mesons. 
The central $V^C_{\bar{q}\bar{q}}$ and the tensor $V^T_{\bar{q}\bar{q}}$ parts of the one-pion exchange potential are taken as
\begin{eqnarray}
\begin{aligned}
V_{\bar{q}\bar{q}}^{C}=&
(\boldsymbol{\sigma}^{(1)} \cdot \boldsymbol{\sigma}^{(2)})
(\boldsymbol{\tau}^{(1)} \cdot \boldsymbol{\tau}^{(2)})
(\frac{g_A}{2f_{\pi}})^2\frac{m_{\pi}^2}{4\pi}\frac{1}{3} \\
&\times
\Big[
\frac{e^{-m_{\pi}r}}{r}
-\frac{e^{-\Lambda r}}{r}
-\frac{\Lambda^2-m_{\pi}^2}{2\Lambda} e^{-\Lambda r}
 \Big],
\end{aligned}
\end{eqnarray}
and
\begin{eqnarray}
\begin{aligned}
V_{\bar{q}\bar{q}}^{T}=&\boldsymbol{S}_{12} (\frac{g_A}{2f_{\pi}})^2\frac{1}{4\pi}\frac{1}{3} \Big[
(3+3m_{\pi}r+{m_{\pi}}^2 r^2) \frac{e^{-m_{\pi}r}}{r^3} 
\\
-&(3+3\Lambda r+{\Lambda}^2 r^2) \frac{e^{-\Lambda r}}{r^3}
+\frac{m_{\pi}^2-\Lambda^2}{2}(1-\Lambda r) \frac{e^{-\Lambda r}}{r} \Big],
\end{aligned}
\end{eqnarray}
where
\begin{eqnarray}
\begin{aligned}
\boldsymbol{S}_{12}=\frac{3(\boldsymbol{\sigma}^{(1)}\cdot \boldsymbol{r})(\boldsymbol{\sigma}^{(2)}\cdot \boldsymbol{r})}{r^2}-\boldsymbol{\sigma}^{(1)}\cdot \boldsymbol{\sigma}^{(2)}.
\end{aligned}
\end{eqnarray}
Here, $m_{\pi}$, $g_A$ and $f_\pi$ are the pion mass, the quark axial coupling and the pion decay constant, respectively, whose values are listed in Table~\ref{parameters}. 
The $\Lambda$ is the cut-off parameter for the short-range part of OPE, which we set $1$ GeV, and later check how the results depend on the choice.

We solve the four-body problem by using the Gaussian expansion method.
The energy and width of the resonances are estimated by the real-scaling method.
The variational wave function $\Psi_{I,JM}$ for states of isospin $I$ and total spin $(J,M)$ is formed
as follows:
\begin{eqnarray}
	\Psi_{I, JM}= & {\cal A} \sum_C \xi_1^{(C)} \sum_{\gamma} B_{\gamma}^{(C)}\eta^{(C)}_I 
  \big[[\chi_{\frac{1}{2}}\chi_{\frac{1}{2}}]_s [\chi_{\frac{1}{2}}
 \chi_{\frac{1}{2}}]_{s'}  \big]_S
 \nonumber\\
	&\times  \big[[\phi^{(C)}_{n\ell}({\bm r}) \phi'^{\,(C)}_{NL}({\bm \rho})]_\Lambda
\psi^{(C)}_{\nu \lambda }(\bm{\lambda}) \big]_{L_{tot}} \bigg]_{JM} ,
\label{total wave function}
\end{eqnarray}
where 
$\xi_1$ stands for the color singlet wave function, 
$\eta$ for the isospin part of light antidiquarks, 
$\chi$ for the spin part of each quark,
and 
$\phi$, $\phi'$, and $\psi$ for the spatial wave functions.
${\cal A}$ denotes the anti-symmetrization among the identical quarks.
The index $\gamma$ denotes collectively all the quantum numbers needed for the expansion, 
$\gamma \equiv \{s, s', S,n,N,\nu,\ell,L,\lambda, L_{tot} \}$.
Energy eigenvalues, $E$, and corresponding expansion coefficients, $B_{\gamma}^{(C)}$, are determined 
by diagonalizing the Hamiltonian matrix computed by the basis functions $\Psi_{I, JM}$. 

\begin{figure}
\centering
\includegraphics[height=5cm]{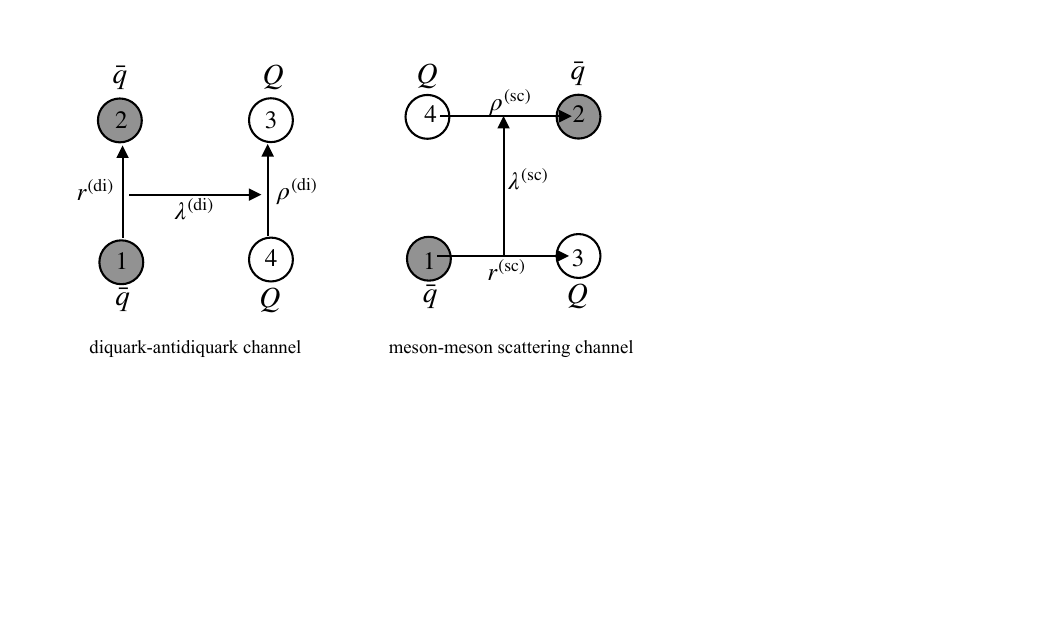}
\caption{ 
The Jacobi coordinates of diquark-antidiquark channel and meson-meson scattering channel for $QQ\bar q \bar q$ tetraquarks. 
} 
\label{fig_jacobi}
\end{figure}

Two sets of Jacobi coordinates corresponding to diquark-antidiquark channel and meson-meson scattering channel shown in Fig.\ref{fig_jacobi} are employed in the present study. The real-scaling technique scales the basis function along $\lambda^{\rm(sc)}$ coordinate of scattering channel by multiplying a scale factor $\alpha$. 
Then the energy eigenvalues of the scattering states decrease as $\alpha$ increases, while the resonant states stay approximately at the resonant energy except for the region of crossing with scattering states.
For more details of real-scaling technique, one can refer to Refs.\cite{realscaling1981, Hiyama:2005cf, Hiyama:2018ukv, Meng:2019fan}.

\section{Results}\label{results}

\begin{figure}
\centering
\includegraphics[height=11cm]{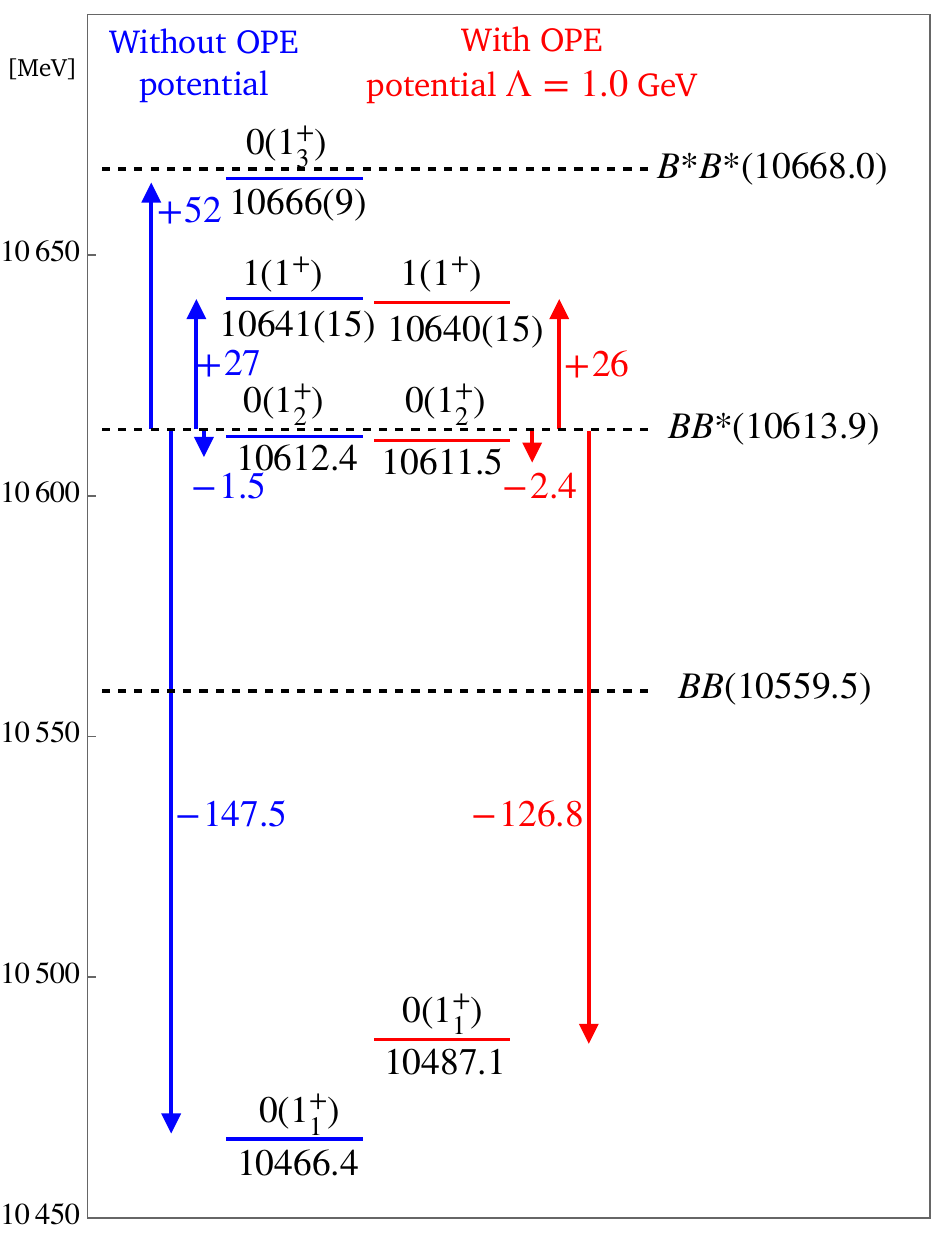}
\caption{ 
Calculated masses of $bb\bar{q}\bar{q}$ bound and resonant states with quantum numbers designated as $I(J^P)$. Blue lines are masses calculated without OPE potential. Red lines are masses calculated including OPE potential for cut-off parameter $\Lambda=1.0$ GeV. The numbers indicate absolute masses as well as the energies measured from the threshold in the unit of MeV. The numbers in parenthesis are the estimated decay width in real-scaling method.
} 
\label{EnenrgyLevelbb}
\end{figure}

\begin{figure}
\centering
\includegraphics[height=4.9cm]{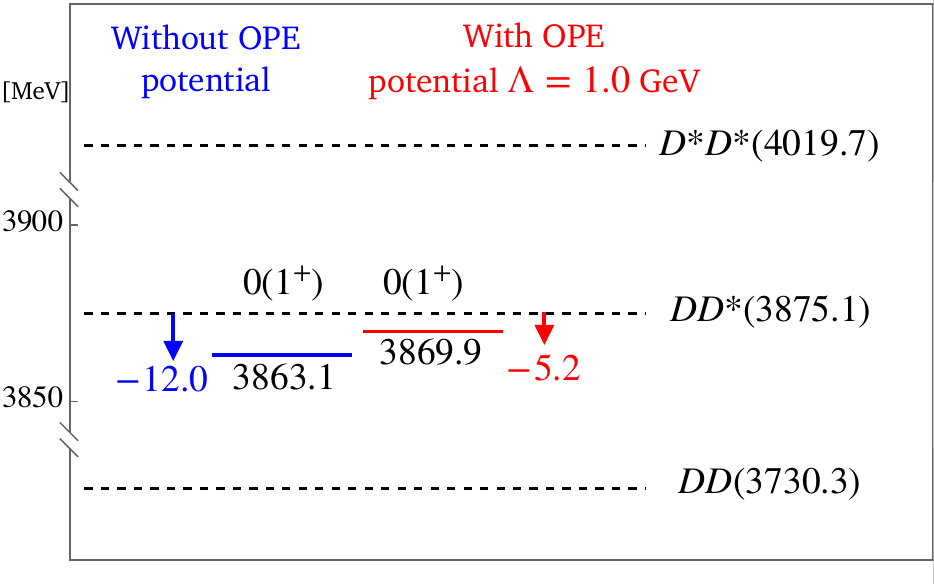}
\caption{ 
Same figure as Fig.\ref{EnenrgyLevelbb} but for $cc\bar{q}\bar{q}$.
} 
\label{EnenrgyLevelcc}
\end{figure}

The obtained energy spectra of positive parity bound and resonant states with quantum numbers designated as $I(J^P)$ for $bb\bar{q}\bar{q}$ and $cc\bar{q}\bar{q}$ are shown in Fig.\ref{EnenrgyLevelbb} and Fig.\ref{EnenrgyLevelcc}, respectively. 
Blue lines are energies calculated without OPE potential. Red lines are energies calculated including OPE potential for cut-off parameter $\Lambda=1.0$ GeV.
In our previous work, we considered isospin $I=0$ states and reported two bound states with $I(J^P)= 0(1^+)$ and two resonant states with $I(J^P)= 0(1^+)$ and $I(J^P)=0(1^-)$ for $bb\bar{q}\bar{q}$. 
We also found a bound state with $I(J^P)= 0(1^+)$ for $cc\bar{q}\bar{q}$.
In this work, we focus on the positive parity states. 
We also study the $I=1$ case and find a resonant state $1(1^+)$ for $bb\bar{q}\bar{q}$ for which the OPE tensor potential between $\bar{q}\bar{q}$ may give a significant contribution.
\footnote{To be noticed, when we apply the new set of parameters, the energies of all bound states and resonant states are slightly shifted and are similar to the results of the previous set of parameters.}
The expectation values of the OPE central and tensor potentials for the each state are listed in Table \ref{Expec}.

\begin{center}
\linespread{1.3}
\begin{table}
\caption{
The energies, $\Delta E=E-E_{th}$, measured from the lowest threshold for the bound and resonant states. Expectation values of the OPE central and tensor potentials are also shown for $\Lambda=1$ GeV.
}
\label{Expec}
\begin{tabular}{p{1.2cm}<{\centering} p{1.3cm}<{\centering} p{0.01cm}<{\centering}p{1.0cm}<{\centering} p{1.0cm}<{\centering}p{1.0cm}<{\centering}  }
\hline\hline
 &  no OPE  && \multicolumn{3}{c}{OPE} \\
\cline{2-2}
\cline{4-6}
  &  $\Delta E$  && $\Delta E$  & $\langle V^C \rangle$ & $\langle V^T \rangle$ \\
\hline
 $bb\bar{q}\bar{q} \ 0(1^+_1)$   &  -147.5  && -126.8 & +20.6  & -0.03   \\
 $bb\bar{q}\bar{q} \ 0(1^+_2)$   &  -1.5	&& -2.4   & -0.4  &  -2.4  \\
 $bb\bar{q}\bar{q} \ 0(1^+_3)$   &  +52     &&  $/$   & $/$  &  $/$ \\
 $bb\bar{q}\bar{q} \ 1(1^+)$     &  +27     && +26    & +1.4   &  -2.5  \\
 $cc\bar{q}\bar{q} \ 0(1^+_1)$   &  -12.0   && -5.2   & +5.6  &  -0.7 \\
\hline\hline
\end{tabular}
\end{table}
\end{center}

One sees that for the $0(1^+)$ ground states of $bb\bar{q}\bar{q}$ and $cc\bar{q}\bar{q}$, the OPE central potential is repulsive, while the tensor contribution is negligibly small. This is expected because $0(1^+_1)$ state is found to be dominated by $S$-wave diquark-antidiquark configuration with isospin $0$ and spin $0$ $\bar{q}\bar{q}$, giving $\langle(\boldsymbol{\sigma}^{(1)} \cdot \boldsymbol{\sigma}^{(2)})(\boldsymbol{\tau}^{(1)} \cdot \boldsymbol{\tau}^{(2)})\rangle =+9$, and  $\langle S_{12}\rangle =0$.
The net OPE contribution is $\sim 21$ MeV for $bb\bar{q}\bar{q}$ and $\sim 7$ MeV for $cc\bar{q}\bar{q}$ for $\Lambda=1$ GeV.

\begin{figure}
\centering
\includegraphics[height=5cm]{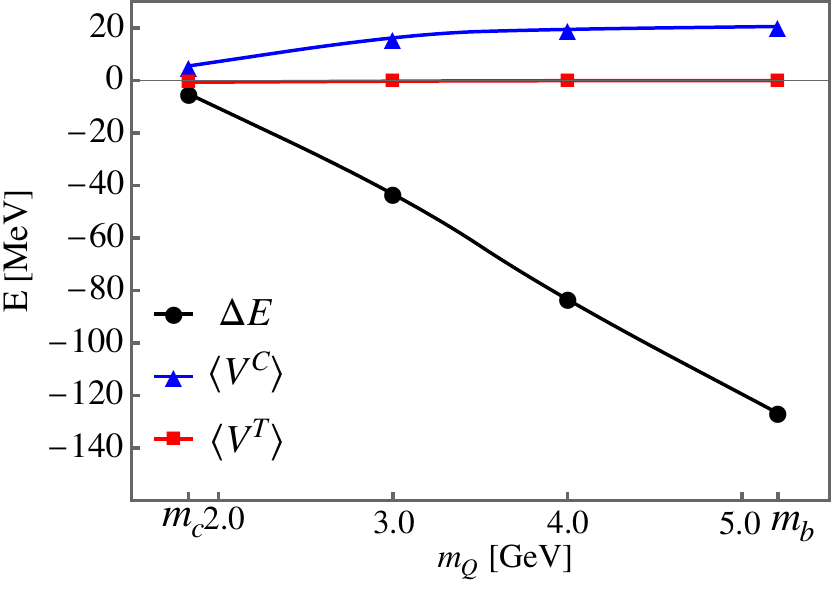}
\caption{ 
The heavy-quark mass dependence of the energy measured from the threshold $\Delta E$ and the expectation values of OPE central potential $\langle V^C \rangle$ and OPE tensor potential $\langle V^T \rangle$ for the deep bound state of $QQ\bar{q}\bar{q} \ \ 0(1^+)$ for $\Lambda=1.0$ GeV. $m_c$ and $m_b$ stand for the mass of $c$ quark and $b$ quark, respectively.
} 
\label{depofmass}
\end{figure}

Fig.\ref{depofmass} plots the energy measured from threshold $\Delta E$, expectation values of OPE central potential $\langle V^C \rangle$ and tensor potential $\langle V^T \rangle$ with $\Lambda=1.0$ GeV as functions of the heavy quark mass. 

\begin{figure}
\centering
\includegraphics[height=10cm]{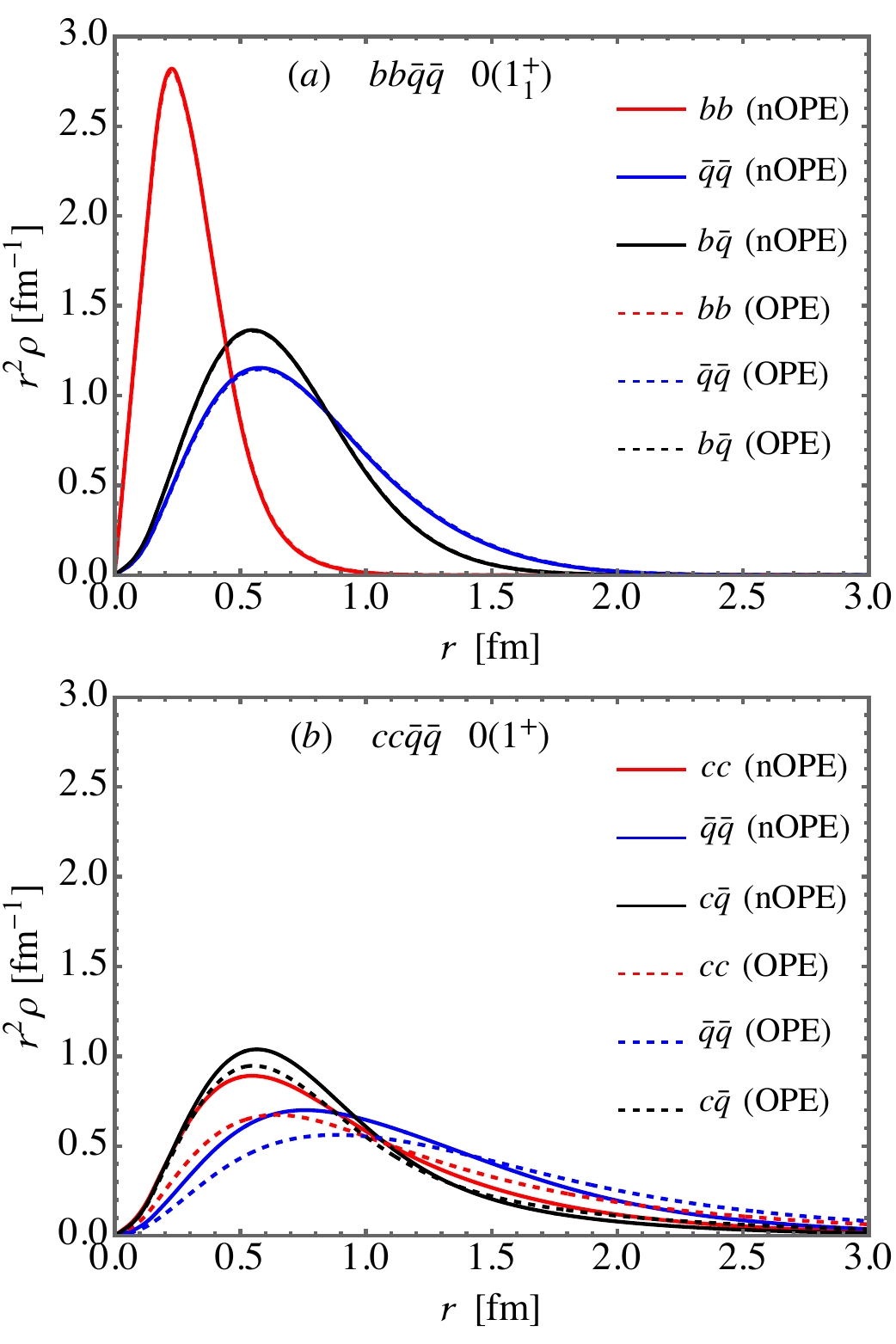}
\caption{ 
Density distributions of different quark pairs for $(a) \ bb\bar{q}\bar{q} \ \ 0(1^+_1)$ and $(b) \ cc\bar{q}\bar{q} \ \ 0(1^+)$. Solid lines are calculated without OPE potential (nOPE) and dashed lines are calculated including OPE potential (OPE) for $\Lambda=1.0$ GeV.} 
\label{abrms}
\end{figure}

\begin{figure}
\centering
\includegraphics[height=10cm]{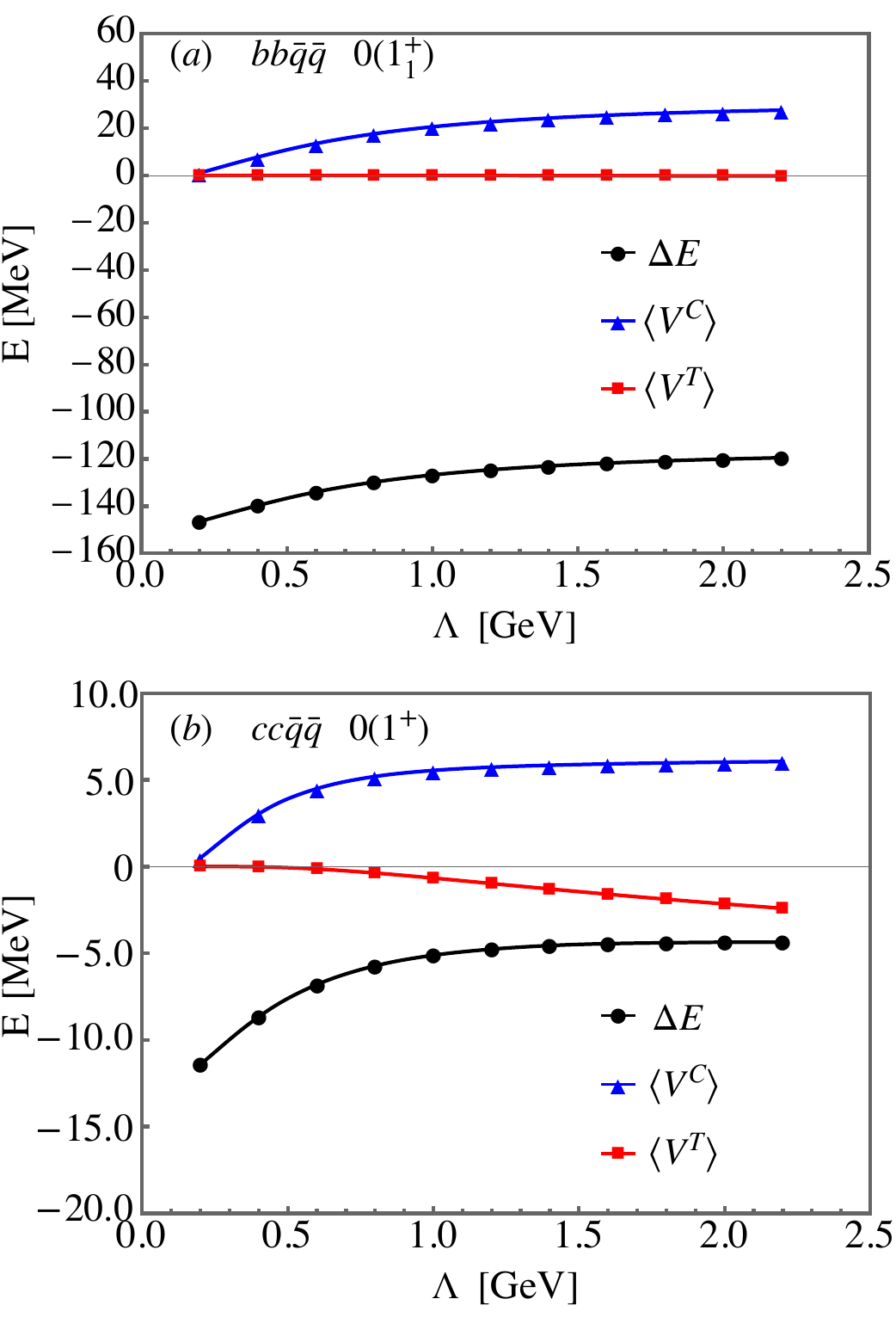}
\caption{ 
The cut-off parameter $\Lambda$ dependence of the energy measured from the threshold $\Delta E$ and the expectation values of OPE central potential $\langle V^C \rangle$ and OPE tensor potential $\langle V^T \rangle$ for $(a) \ bb\bar{q}\bar{q} \ \ 0(1^+_1)$ and $(b) \ cc\bar{q}\bar{q} \ \ 0(1^+)$.
} 
\label{ablam}
\end{figure}

In order to investigate the effect of OPE potential to their dynamical structure, we calculate the two-body density distributions of quark pairs.
Fig.\ref{abrms}(a) and Fig.\ref{abrms}(b) show the distributions as functions of distances between different quark pairs with or without OPE potential in $bb\bar{q}\bar{q}$ $0(1^+_1)$ and $cc\bar{q}\bar{q}$ $0(1^+)$, respectively.
One sees that the quark distribution of the compact tetraquark state, $bb\bar{q}\bar{q}$ $0(1^+_1)$, barely changes by the OPE potential.
In contrast, OPE repels all the quark pairs in the molecular tetraquark state, $cc\bar{q}\bar{q}$ $0(1^+)$.
This result indicates that the effect of OPE potential to the spatial structure of the compact tetraquark is much smaller than that to the molecular tetraquark.

We also study the dependence on the cut-off parameter, $\Lambda$. 
Fig.\ref{ablam}(a) and Fig.\ref{ablam}(b) show the $\Lambda$ dependences of $\Delta E$, $\langle V^C \rangle$ and $\langle V^T \rangle$ for $bb\bar{q}\bar{q} \ 0(1^+_1)$ and $cc\bar{q}\bar{q} \ 0(1^+)$.
One sees that for $bb\bar{q}\bar{q}$, with the increase of $\Lambda$, contribution of the central OPE increases and saturates around 6 MeV, while the contribution of the tensor force is always negative and grows monotonically.
The behavior of $\Delta E$ is governed by the central potential since the tensor contribution is small. 
In the case of $cc\bar{q}\bar{q} \ 0(1^+)$, the situation is similar to the $bb\bar{q}\bar{q}$ case for central potential and $\Delta E$, while tensor contribution increases slowly. This is because the molecular component is larger in $cc\bar{q}\bar{q}$ than in $bb\bar{q}\bar{q}$, where the tensor force becomes more effective.

\begin{figure}
\centering
\includegraphics[height=5cm]{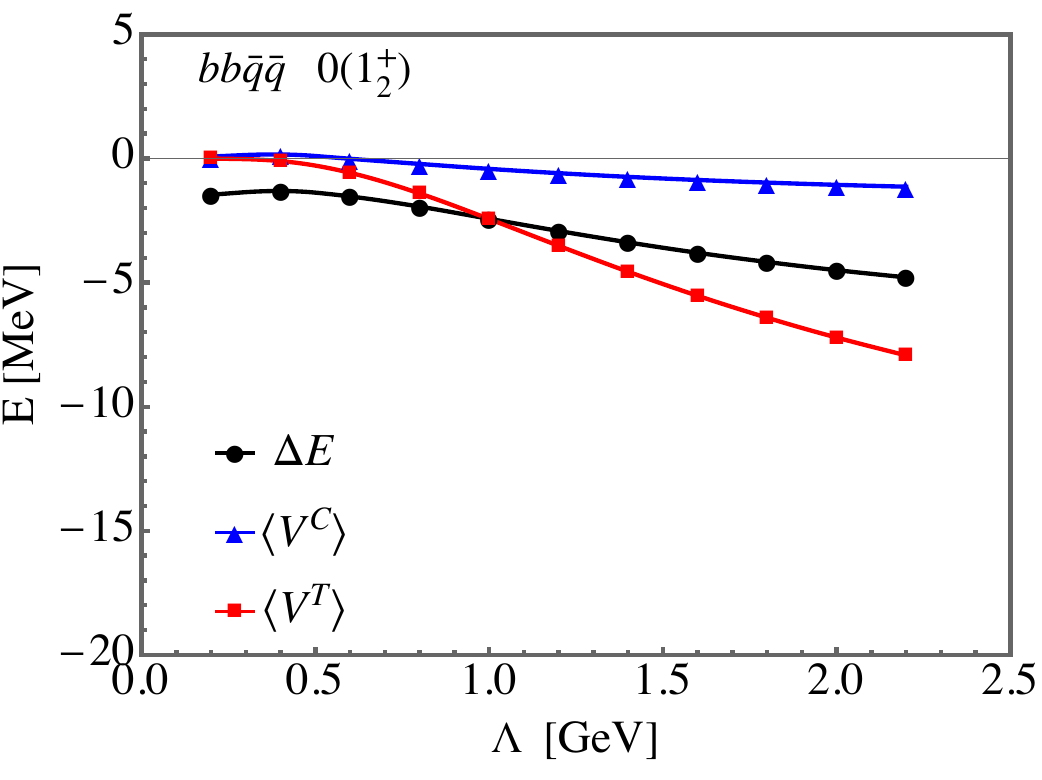}
\caption{ 
The cut-off parameter $\Lambda$ dependence of the energy measured from the threshold $\Delta E$ and the expectation values of OPE central potential $\langle V^C \rangle$ and OPE tensor potential $\langle V^T \rangle$ for $\ bb\bar{q}\bar{q} \ \ 0(1^+_2)$.
} 
\label{Elam_bb_shallow}
\end{figure}

\begin{figure}
\centering
\includegraphics[height=5cm]{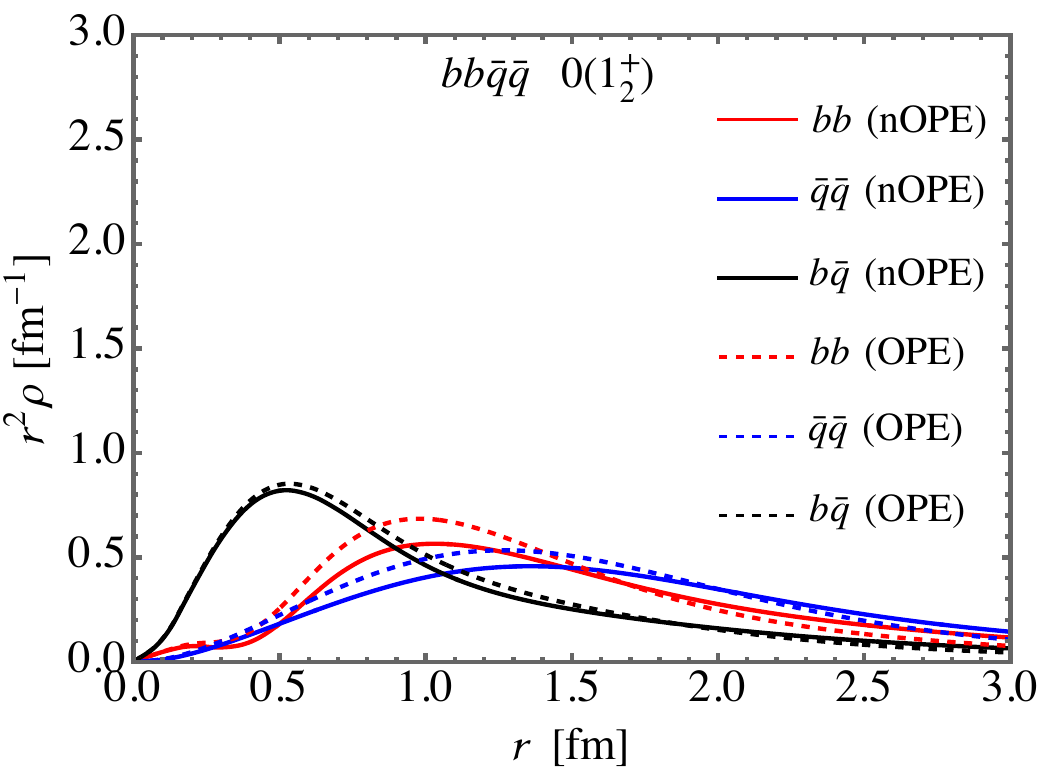}
\caption{ 
Density distributions of different quark pairs for $\ bb\bar{q}\bar{q} \ \ 1(1^+)$. Solid lines are calculated without OPE potential (nOPE) and dashed lines are calculated including OPE potential (OPE) for $\Lambda=1.0$ GeV.} 
\label{rms_bbiso0shallow}
\end{figure}

Next we discuss the excited states, $0(1^+_2)$ and $0(1^+_3)$ in $bb\bar{q}\bar{q}$.
The $bb\bar{q}\bar{q} \ 0(1^+_2)$ state is shifted down by $\sim$ 1 MeV when OPE potential is included with $\Lambda=1.0$ GeV. The contribution of tensor potential and central potential are -2.4 MeV and -0.4 MeV, respectively. 
This state is found to be dominated by $BB^*$ configuration which contains both spin $1$ $\bar{q}\bar{q}$ component and spin $0$ $\bar{q}\bar{q}$ component.
We conjecture that spin $1$ $\bar{q}\bar{q}$ component is dominant, leading to the negative tensor contribution and negative central contributions.
Fig.\ref{Elam_bb_shallow} shows the cut-off parameter $\Lambda$ dependence of $\Delta E$, $\langle V^C \rangle$ and $\langle V^T \rangle$. Fig.\ref{rms_bbiso0shallow} shows the density distributions of different quark pairs. One can see that as a molecular tetraquark state, $bb\bar{q}\bar{q} \ 0(1^+_2)$ becomes more compact when the OPE potential is included. This also shows that OPE has a relative large effect to a molecular tetraquark state.

The $bb \bar{q}\bar{q}$ $0(1_3^+)$ state disappears when OPE is turned on.
It is possible that it is shifted up and is mixed with the $B^*B^*$ scattering states. This is reasonable because $0(1^+_3)$ state is a mix of $2S$-excited $bb$ diquark-antidiquark configuration and $B^*B^*$ which has been shown in our previous paper. Similar to $0(1^+_1)$, the repulsive central potential pushes it above the $B^*B^*$ threshold.

We note that another set of narrow $bb\bar{q}\bar{q}$ resonances is found with isospin $I=0$, $J^P=0^+$, $1^+$ and $2^+$ (not shown in Fig.\ref{EnenrgyLevelbb}). It is located at 350 MeV above the $BB^*$ threshold with decay width $\Gamma \sim \rm 10 MeV$ when the OPE potential is not included.
The main component is given by the following configuration: the relative angular momentum $l$ for $\bar{q}\bar{q}$ is 0, $L$ for $bb$ is 1 and $\lambda$ for $\bar{q}\bar{q}-bb$ is 1.
By including the OPE potential, it is shifted up by $\sim 15 \rm MeV$ which is almost all attributed to OPE central potential. This result is consistent with those of $bb\bar{q}\bar{q}$ $0(1^+_1)$ state and $cc\bar{q}\bar{q}$ $0(1^+)$ state, both of which have S-wave $\bar{q}\bar{q}$ in dominating component. 

\begin{figure}
\centering
\includegraphics[height=5cm]{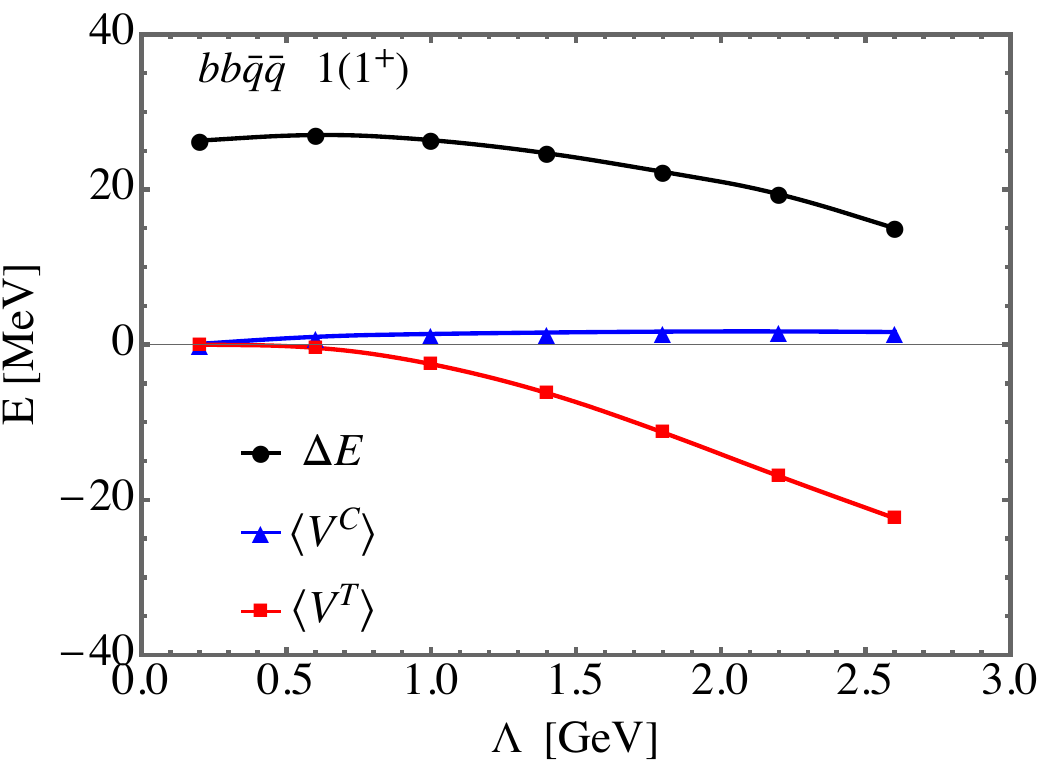}
\caption{ 
The cut-off parameter $\Lambda$ dependence of the energy measured from the threshold $\Delta E$ and the expectation values of OPE central potential $\langle V^C \rangle$ and OPE tensor potential $\langle V^T \rangle$ for $\ bb\bar{q}\bar{q} \ \ 1(1^+)$.
} 
\label{Elam_bbiso1}
\end{figure}

\begin{figure}
\centering
\includegraphics[height=5cm]{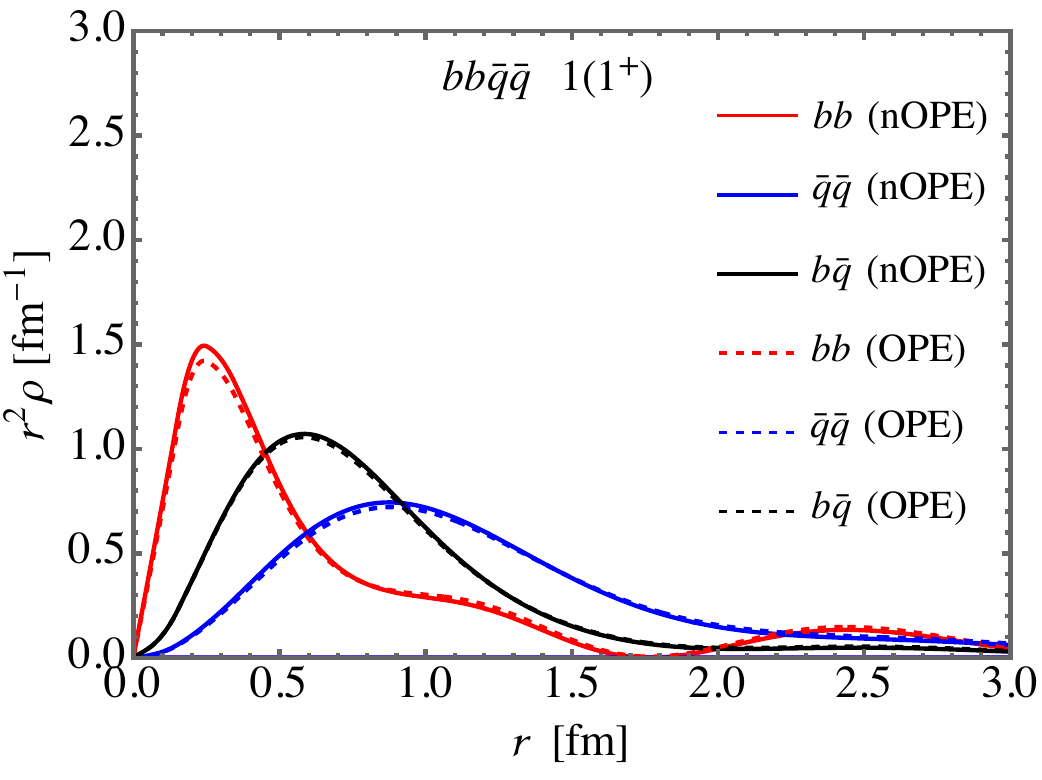}
\caption{ 
Density distributions of different quark pairs for $\ bb\bar{q}\bar{q} \ \ 1(1^+)$. Solid lines are calculated without OPE potential (nOPE) and dashed lines are calculated including OPE potential (OPE) for $\Lambda=1.0$ GeV.} 
\label{rms_bbiso1}
\end{figure}

Finally we discuss the isospin $1$ state, $bb\bar{q}\bar{q}$ $1(1^+)$, that is newly found in the present calculation. When the OPE potential is not included, it is located at 27 MeV above the $BB^*$ threshold with decay width 15 MeV.
It is dominated by the $S$-wave diquark-antidiquark configuration where the $\bar{q}\bar{q}$ $I(J^P)=1(1^+)$ is bound to the heavy vector diquark $bb$. Thus the tensor and central potentials are repulsive and attractive, respectively. When $\Lambda=1.0$ GeV, $\langle V^T \rangle$ is -2.5 MeV and $\langle V^C \rangle$ is +1.4 MeV.
The net contribution is $\sim$ -1 MeV.
The $\Lambda$ dependence of OPE potential contributions for $bb\bar{q}\bar{q}$ $1(1^+)$ state is different from that for isospin 0 state.
Fig.\ref{Elam_bbiso1} shows the cut-off parameter $\Lambda$ dependence of $\Delta E$, $\langle V^C \rangle$ and $\langle V^T \rangle$ for $1(1^+)$ state.
When $\Lambda$ increases, unlike the isospin 0 states, the $\langle V^C \rangle$ remains repulsive and increases very slowly while the $\langle V^T \rangle$ is attractive and increases rapidly.  
The density distributions shown in Fig.\ref{rms_bbiso1} indicate that the $1(1^+)$ state has compact structure and OPE potential barely changes its structure.

The vector $\bar{q}\bar{q}$ and vector $bb$ diquarks allow
tetraquarks with $I(J^P)=1(0^+)$ and $1(2^+)$ also.  
For $1(0^+)$ state, it is strongly coupled to the $BB$ scattering state and does not appear as a resonant state. Previously, for $1(2^+)$ state, we reported that this state was located about 3 MeV below the $B^*B^*$ threshold \cite{Meng:2020knc}. However, this bound state is shifted up and melts into the $B^*B^*$ scattering state in the present results due to the slightly weaker quark-quark potential.

\section{Summary}\label{summary}

In this work, we have studied the spectra of the $bb\bar{q}\bar{q}$ and $cc\bar{q}\bar{q}$ tetraquarks with one-pion exchange potential between two light antiquarks taken into consideration. 
The effect of OPE is found not so strong to change the qualitative behaviors of the tetraquark spectrum.
Compared with the results without OPE potential, the binding energies of $bb\bar{q}\bar{q} \ 0(1^+_1)$ and $cc\bar{q}\bar{q} \ 0(1^+)$ bound states become smaller as shown in Fig.~\ref{EnenrgyLevelbb} and Fig.~\ref{EnenrgyLevelcc} because of the repulsive central force between the $I=0$ scalar $\bar{q}\bar{q}$. 
The suppression of the binding energy seems consistent with the observation of $T_{cc}$ just below the threshold of $DD^*$ by the LHCb.
We have also found a new $bb\bar{q}\bar{q}$ resonant state with $I(J^P)=1(1^+)$. Its main component is a $I=1$ vector $\bar{q}\bar{q}$ bound to a vector $bb$ which leads to a repulsive central force and attractive tensor force between $\bar{q}\bar{q}$. They tend to cancel with each other and leave the small energy difference when OPE potential is included.
Our results indicate that the effect of OPE potential to the spatial structure of the molecular tetraquark is much larger than that to the compact tetraquark.

\section*{Acknowledgments}
This work is supported by the National Natural Science Foundation of China, Grant No. 12275129 and the Fundamental Research Funds for the Central Universities, Grant No. 020414380209 (QM and CX), and the Japanese Grant-in-Aid for Scientific Research, Grant Nos. 21H04478 and 18H05407 (AH), 20K03959, 21H00132 and 23K03427 (MO).  

\printcredits

\bibliographystyle{elsarticle-num}

\bibliography{cas-refs}

\begin{thebibliography}{10}
\expandafter\ifx\csname url\endcsname\relax
  \def\url#1{\texttt{#1}}\fi
\expandafter\ifx\csname urlprefix\endcsname\relax\def\urlprefix{URL }\fi
\expandafter\ifx\csname href\endcsname\relax
  \def\href#1#2{#2} \def\path#1{#1}\fi

\bibitem{LHCb:2021vvq}
R.~Aaij, et~al., {Observation of an exotic narrow doubly charmed tetraquark},
  Nature Phys. 18~(7) (2022) 751--754.
\newblock \href {http://arxiv.org/abs/2109.01038} {\path{arXiv:2109.01038}},
  \href {http://dx.doi.org/10.1038/s41567-022-01614-y}
  {\path{doi:10.1038/s41567-022-01614-y}}.

\bibitem{Belle:2003nnu}
S.~K. Choi, et~al., {Observation of a narrow charmonium-like state in exclusive
  $B^\pm \to K^\pm \pi^+ \pi^- J/\psi$ decays}, Phys. Rev. Lett. 91 (2003)
  262001.
\newblock \href {http://arxiv.org/abs/hep-ex/0309032}
  {\path{arXiv:hep-ex/0309032}}, \href
  {http://dx.doi.org/10.1103/PhysRevLett.91.262001}
  {\path{doi:10.1103/PhysRevLett.91.262001}}.

\bibitem{LHCb:2015yax}
R.~Aaij, et~al., {Observation of $J/\psi p$ Resonances Consistent with
  Pentaquark States in $\Lambda_b^0 \to J/\psi K^- p$ Decays}, Phys. Rev. Lett.
  115 (2015) 072001.
\newblock \href {http://arxiv.org/abs/1507.03414} {\path{arXiv:1507.03414}},
  \href {http://dx.doi.org/10.1103/PhysRevLett.115.072001}
  {\path{doi:10.1103/PhysRevLett.115.072001}}.

\bibitem{Ballot:1983iv}
J.~l. Ballot, J.~M. Richard, {FOUR QUARK STATES IN ADDITIVE POTENTIALS}, Phys.
  Lett. B 123 (1983) 449--451.
\newblock \href {http://dx.doi.org/10.1016/0370-2693(83)90991-7}
  {\path{doi:10.1016/0370-2693(83)90991-7}}.

\bibitem{Zouzou:1986qh}
S.~Zouzou, B.~Silvestre-Brac, C.~Gignoux, J.~M. Richard, {FOUR QUARK BOUND
  STATES}, Z. Phys. C 30 (1986) 457.
\newblock \href {http://dx.doi.org/10.1007/BF01557611}
  {\path{doi:10.1007/BF01557611}}.

\bibitem{Carlson:1987hh}
J.~Carlson, L.~Heller, J.~A. Tjon, {Stability of Dimesons}, Phys. Rev. D 37
  (1988) 744.
\newblock \href {http://dx.doi.org/10.1103/PhysRevD.37.744}
  {\path{doi:10.1103/PhysRevD.37.744}}.

\bibitem{Silvestre-Brac:1993zem}
B.~Silvestre-Brac, C.~Semay, {Systematics of L = 0 q-2 anti-q-2 systems}, Z.
  Phys. C 57 (1993) 273--282.
\newblock \href {http://dx.doi.org/10.1007/BF01565058}
  {\path{doi:10.1007/BF01565058}}.

\bibitem{Pepin:1996id}
S.~Pepin, F.~Stancu, M.~Genovese, J.~M. Richard, {Tetraquarks with color blind
  forces in chiral quark models}, Phys. Lett. B 393 (1997) 119--123.
\newblock \href {http://arxiv.org/abs/hep-ph/9609348}
  {\path{arXiv:hep-ph/9609348}}, \href
  {http://dx.doi.org/10.1016/S0370-2693(96)01597-3}
  {\path{doi:10.1016/S0370-2693(96)01597-3}}.

\bibitem{Gelman:2002wf}
B.~A. Gelman, S.~Nussinov, {Does a narrow tetraquark cc anti-u anti-d state
  exist?}, Phys. Lett. B 551 (2003) 296--304.
\newblock \href {http://arxiv.org/abs/hep-ph/0209095}
  {\path{arXiv:hep-ph/0209095}}, \href
  {http://dx.doi.org/10.1016/S0370-2693(02)03069-1}
  {\path{doi:10.1016/S0370-2693(02)03069-1}}.

\bibitem{Vijande:2003ki}
J.~Vijande, F.~Fernandez, A.~Valcarce, B.~Silvestre-Brac, {Tetraquarks in a
  chiral constituent quark model}, Eur. Phys. J. A 19 (2004) 383.
\newblock \href {http://arxiv.org/abs/hep-ph/0310007}
  {\path{arXiv:hep-ph/0310007}}, \href
  {http://dx.doi.org/10.1140/epja/i2003-10128-9}
  {\path{doi:10.1140/epja/i2003-10128-9}}.

\bibitem{Ebert:2007rn}
D.~Ebert, R.~N. Faustov, V.~O. Galkin, W.~Lucha, {Masses of tetraquarks with
  two heavy quarks in the relativistic quark model}, Phys. Rev. D 76 (2007)
  114015.
\newblock \href {http://arxiv.org/abs/0706.3853} {\path{arXiv:0706.3853}},
  \href {http://dx.doi.org/10.1103/PhysRevD.76.114015}
  {\path{doi:10.1103/PhysRevD.76.114015}}.

\bibitem{Yang:2009zzp}
Y.~Yang, C.~Deng, J.~Ping, T.~Goldman, {S-wave Q Q anti-q anti-q state in the
  constituent quark model}, Phys. Rev. D 80 (2009) 114023.
\newblock \href {http://dx.doi.org/10.1103/PhysRevD.80.114023}
  {\path{doi:10.1103/PhysRevD.80.114023}}.

\bibitem{Ikeda:2013vwa}
Y.~Ikeda, B.~Charron, S.~Aoki, T.~Doi, T.~Hatsuda, T.~Inoue, N.~Ishii,
  K.~Murano, H.~Nemura, K.~Sasaki, {Charmed tetraquarks $T_{cc}$ and $T_{cs}$
  from dynamical lattice QCD simulations}, Phys. Lett. B 729 (2014) 85--90.
\newblock \href {http://arxiv.org/abs/1311.6214} {\path{arXiv:1311.6214}},
  \href {http://dx.doi.org/10.1016/j.physletb.2014.01.002}
  {\path{doi:10.1016/j.physletb.2014.01.002}}.

\bibitem{Luo:2017eub}
S.-Q. Luo, K.~Chen, X.~Liu, Y.-R. Liu, S.-L. Zhu, {Exotic tetraquark states
  with the $qq\bar{Q}\bar{Q}$ configuration}, Eur. Phys. J. C 77~(10) (2017)
  709.
\newblock \href {http://arxiv.org/abs/1707.01180} {\path{arXiv:1707.01180}},
  \href {http://dx.doi.org/10.1140/epjc/s10052-017-5297-4}
  {\path{doi:10.1140/epjc/s10052-017-5297-4}}.

\bibitem{Karliner:2017qjm}
M.~Karliner, J.~L. Rosner, {Discovery of doubly-charmed $\Xi_{cc}$ baryon
  implies a stable ($b b \bar{u} \bar{d}$) tetraquark}, Phys. Rev. Lett.
  119~(20) (2017) 202001.
\newblock \href {http://arxiv.org/abs/1707.07666} {\path{arXiv:1707.07666}},
  \href {http://dx.doi.org/10.1103/PhysRevLett.119.202001}
  {\path{doi:10.1103/PhysRevLett.119.202001}}.

\bibitem{Eichten:2017ffp}
E.~J. Eichten, C.~Quigg, {Heavy-quark symmetry implies stable heavy tetraquark
  mesons $Q_iQ_j \bar q_k \bar q_l$}, Phys. Rev. Lett. 119~(20) (2017) 202002.
\newblock \href {http://arxiv.org/abs/1707.09575} {\path{arXiv:1707.09575}},
  \href {http://dx.doi.org/10.1103/PhysRevLett.119.202002}
  {\path{doi:10.1103/PhysRevLett.119.202002}}.

\bibitem{Liu:2019stu}
M.-Z. Liu, T.-W. Wu, M.~Pavon~Valderrama, J.-J. Xie, L.-S. Geng, {Heavy-quark
  spin and flavor symmetry partners of the X(3872) revisited: What can we learn
  from the one boson exchange model?}, Phys. Rev. D 99~(9) (2019) 094018.
\newblock \href {http://arxiv.org/abs/1902.03044} {\path{arXiv:1902.03044}},
  \href {http://dx.doi.org/10.1103/PhysRevD.99.094018}
  {\path{doi:10.1103/PhysRevD.99.094018}}.

\bibitem{Junnarkar:2018twb}
P.~Junnarkar, N.~Mathur, M.~Padmanath, {Study of doubly heavy tetraquarks in
  Lattice QCD}, Phys. Rev. D 99~(3) (2019) 034507.
\newblock \href {http://arxiv.org/abs/1810.12285} {\path{arXiv:1810.12285}},
  \href {http://dx.doi.org/10.1103/PhysRevD.99.034507}
  {\path{doi:10.1103/PhysRevD.99.034507}}.

\bibitem{Francis:2018jyb}
A.~Francis, R.~J. Hudspith, R.~Lewis, K.~Maltman, {Evidence for charm-bottom
  tetraquarks and the mass dependence of heavy-light tetraquark states from
  lattice QCD}, Phys. Rev. D 99~(5) (2019) 054505.
\newblock \href {http://arxiv.org/abs/1810.10550} {\path{arXiv:1810.10550}},
  \href {http://dx.doi.org/10.1103/PhysRevD.99.054505}
  {\path{doi:10.1103/PhysRevD.99.054505}}.

\bibitem{Lu:2020rog}
Q.-F. L\"u, D.-Y. Chen, Y.-B. Dong, {Masses of doubly heavy tetraquarks
  $T_{QQ^\prime}$ in a relativized quark model}, Phys. Rev. D 102~(3) (2020)
  034012.
\newblock \href {http://arxiv.org/abs/2006.08087} {\path{arXiv:2006.08087}},
  \href {http://dx.doi.org/10.1103/PhysRevD.102.034012}
  {\path{doi:10.1103/PhysRevD.102.034012}}.

\bibitem{Deng:2018kly}
C.~Deng, H.~Chen, J.~Ping, {Systematical investigation on the stability of
  doubly heavy tetraquark states}, Eur. Phys. J. A 56~(1) (2020) 9.
\newblock \href {http://arxiv.org/abs/1811.06462} {\path{arXiv:1811.06462}},
  \href {http://dx.doi.org/10.1140/epja/s10050-019-00012-y}
  {\path{doi:10.1140/epja/s10050-019-00012-y}}.

\bibitem{Cheng:2020wxa}
J.-B. Cheng, S.-Y. Li, Y.-R. Liu, Z.-G. Si, T.~Yao, {Double-heavy tetraquark
  states with heavy diquark-antiquark symmetry}, Chin. Phys. C 45~(4) (2021)
  043102.
\newblock \href {http://arxiv.org/abs/2008.00737} {\path{arXiv:2008.00737}},
  \href {http://dx.doi.org/10.1088/1674-1137/abde2f}
  {\path{doi:10.1088/1674-1137/abde2f}}.

\bibitem{Braaten:2020nwp}
E.~Braaten, L.-P. He, A.~Mohapatra, {Masses of doubly heavy tetraquarks with
  error bars}, Phys. Rev. D 103~(1) (2021) 016001.
\newblock \href {http://arxiv.org/abs/2006.08650} {\path{arXiv:2006.08650}},
  \href {http://dx.doi.org/10.1103/PhysRevD.103.016001}
  {\path{doi:10.1103/PhysRevD.103.016001}}.

\bibitem{Dong:2021bvy}
X.-K. Dong, F.-K. Guo, B.-S. Zou, {A survey of heavy\textendash{}heavy hadronic
  molecules}, Commun. Theor. Phys. 73~(12) (2021) 125201.
\newblock \href {http://arxiv.org/abs/2108.02673} {\path{arXiv:2108.02673}},
  \href {http://dx.doi.org/10.1088/1572-9494/ac27a2}
  {\path{doi:10.1088/1572-9494/ac27a2}}.

\bibitem{Chen:2021vhg}
R.~Chen, Q.~Huang, X.~Liu, S.-L. Zhu, {Predicting another doubly charmed
  molecular resonance Tcc'+ (3876)}, Phys. Rev. D 104~(11) (2021) 114042.
\newblock \href {http://arxiv.org/abs/2108.01911} {\path{arXiv:2108.01911}},
  \href {http://dx.doi.org/10.1103/PhysRevD.104.114042}
  {\path{doi:10.1103/PhysRevD.104.114042}}.

\bibitem{Weng:2021hje}
X.-Z. Weng, W.-Z. Deng, S.-L. Zhu, {Doubly heavy tetraquarks in an extended
  chromomagnetic model *}, Chin. Phys. C 46~(1) (2022) 013102.
\newblock \href {http://arxiv.org/abs/2108.07242} {\path{arXiv:2108.07242}},
  \href {http://dx.doi.org/10.1088/1674-1137/ac2ed0}
  {\path{doi:10.1088/1674-1137/ac2ed0}}.

\bibitem{Chen:2021cfl}
K.~Chen, R.~Chen, L.~Meng, B.~Wang, S.-L. Zhu, {Systematics of the heavy flavor
  hadronic molecules}, Eur. Phys. J. C 82~(7) (2022) 581.
\newblock \href {http://arxiv.org/abs/2109.13057} {\path{arXiv:2109.13057}},
  \href {http://dx.doi.org/10.1140/epjc/s10052-022-10540-5}
  {\path{doi:10.1140/epjc/s10052-022-10540-5}}.

\bibitem{Deng:2021gnb}
C.~Deng, S.-L. Zhu, {Tcc+ and its partners}, Phys. Rev. D 105~(5) (2022)
  054015.
\newblock \href {http://arxiv.org/abs/2112.12472} {\path{arXiv:2112.12472}},
  \href {http://dx.doi.org/10.1103/PhysRevD.105.054015}
  {\path{doi:10.1103/PhysRevD.105.054015}}.

\bibitem{Padmanath:2022cvl}
M.~Padmanath, S.~Prelovsek, {Signature of a Doubly Charm Tetraquark Pole in DD*
  Scattering on the Lattice}, Phys. Rev. Lett. 129~(3) (2022) 032002.
\newblock \href {http://arxiv.org/abs/2202.10110} {\path{arXiv:2202.10110}},
  \href {http://dx.doi.org/10.1103/PhysRevLett.129.032002}
  {\path{doi:10.1103/PhysRevLett.129.032002}}.

\bibitem{Kim:2022mpa}
Y.~Kim, M.~Oka, K.~Suzuki, {Doubly heavy tetraquarks in a chiral-diquark
  picture}, Phys. Rev. D 105~(7) (2022) 074021.
\newblock \href {http://arxiv.org/abs/2202.06520} {\path{arXiv:2202.06520}},
  \href {http://dx.doi.org/10.1103/PhysRevD.105.074021}
  {\path{doi:10.1103/PhysRevD.105.074021}}.

\bibitem{Cheng:2022qcm}
J.-B. Cheng, Z.-Y. Lin, S.-L. Zhu, {Double-charm tetraquark under the complex
  scaling method}, Phys. Rev. D 106~(1) (2022) 016012.
\newblock \href {http://arxiv.org/abs/2205.13354} {\path{arXiv:2205.13354}},
  \href {http://dx.doi.org/10.1103/PhysRevD.106.016012}
  {\path{doi:10.1103/PhysRevD.106.016012}}.

\bibitem{Meng:2020knc}
Q.~Meng, E.~Hiyama, A.~Hosaka, M.~Oka, P.~Gubler, K.~U. Can, T.~T. Takahashi,
  H.~S. Zong, {Stable double-heavy tetraquarks: spectrum and structure}, Phys.
  Lett. B 814 (2021) 136095.
\newblock \href {http://arxiv.org/abs/2009.14493} {\path{arXiv:2009.14493}},
  \href {http://dx.doi.org/10.1016/j.physletb.2021.136095}
  {\path{doi:10.1016/j.physletb.2021.136095}}.

\bibitem{Meng:2021yjr}
Q.~Meng, M.~Harada, E.~Hiyama, A.~Hosaka, M.~Oka, Doubly heavy tetraquark
  resonant states, Phys. Lett. B 824 (2022) 136800.

\bibitem{SilvestreBrac:1996bg}
B.~Silvestre-Brac, {Spectrum and static properties of heavy baryons}, Few Body
  Syst. 20 (1996) 1--25.
\newblock \href {http://dx.doi.org/10.1007/s006010050028}
  {\path{doi:10.1007/s006010050028}}.

\bibitem{realscaling1981}
J.~Simons, Resonance state lifetimes from stabilization graphs, The Journal of
  Chemical Physics 75~(5) (1981) 2465--2467.

\bibitem{Hiyama:2005cf}
E.~Hiyama, M.~Kamimura, A.~Hosaka, H.~Toki, M.~Yahiro, {Five-body calculation
  of resonance and scattering states of pentaquark system}, Phys. Lett. B 633
  (2006) 237--244.
\newblock \href {http://arxiv.org/abs/hep-ph/0507105}
  {\path{arXiv:hep-ph/0507105}}, \href
  {http://dx.doi.org/10.1016/j.physletb.2005.11.086}
  {\path{doi:10.1016/j.physletb.2005.11.086}}.

\bibitem{Hiyama:2018ukv}
E.~Hiyama, A.~Hosaka, M.~Oka, J.-M. Richard, {Quark model estimate of
  hidden-charm pentaquark resonances}, Phys. Rev. C 98~(4) (2018) 045208.
\newblock \href {http://arxiv.org/abs/1803.11369} {\path{arXiv:1803.11369}},
  \href {http://dx.doi.org/10.1103/PhysRevC.98.045208}
  {\path{doi:10.1103/PhysRevC.98.045208}}.

\bibitem{Meng:2019fan}
Q.~Meng, E.~Hiyama, K.~U. Can, P.~Gubler, M.~Oka, A.~Hosaka, H.~Zong, {Compact
  $sssc\bar{c}$ pentaquark states predicted by a quark model}, Phys. Lett. B
  798 (2019) 135028.
\newblock \href {http://arxiv.org/abs/1907.00144} {\path{arXiv:1907.00144}},
  \href {http://dx.doi.org/10.1016/j.physletb.2019.135028}
  {\path{doi:10.1016/j.physletb.2019.135028}}.

\end{thebibliography}

\end{document}